\documentclass[aps,prl,superscriptaddress,twocolumn]{revtex4-1}

\usepackage{siunitx}
\usepackage{amsmath}
\usepackage{graphicx}
\usepackage{longtable}

\usepackage[english]{babel}
\usepackage{blindtext}

\usepackage{soul}
\usepackage{color}

\begin{document}


\title{Transition strength measurements to guide magic wavelength selection in optically trapped molecules}



\author{K. H. Leung}
\affiliation{Department of Physics, Columbia University, 538 West 120th Street, New York, NY 10027-5255, USA}
\author{I. Majewska}
\affiliation{Quantum Chemistry Laboratory, Department of Chemistry,
University of Warsaw, Pasteura 1, 02-093 Warsaw, Poland}
\author{H. Bekker}
\author{C.-H. Lee}
\author{E. Tiberi}
\author{S. S. Kondov}
\altaffiliation{Present address: Atom Computing, Inc., 918 Parker Street, Berkeley, CA 94710, USA.}
\affiliation{Department of Physics, Columbia University, 538 West 120th Street, New York, NY 10027-5255, USA}
\author{R. Moszynski}
\affiliation{Quantum Chemistry Laboratory, Department of Chemistry,
University of Warsaw, Pasteura 1, 02-093 Warsaw, Poland}
\author{T. Zelevinsky}
\email{tanya.zelevinsky@columbia.edu}
\affiliation{Department of Physics, Columbia University, 538 West 120th Street, New York, NY 10027-5255, USA}


\date{\today}

\begin{abstract}
Optical trapping of molecules with long coherence times is crucial for many protocols in quantum information and metrology. However, the factors that limit the lifetimes of the trapped molecules remain elusive and require improved understanding of the underlying molecular structure. Here we show that measurements of vibronic line strengths in weakly and deeply bound $^{88}\rm{Sr}_2$ molecules, combined with \textit{ab initio} calculations, allow for unambiguous identification of vibrational quantum numbers. This, in turn, enables the construction of refined excited potential energy curves that inform the selection of magic wavelengths which facilitate long vibrational coherence.  We demonstrate Rabi oscillations between far-separated vibrational states that persist for nearly 100 ms.
\end{abstract} 


\maketitle

The interplay of light and matter has enabled major strides in creating and controlling ultracold atoms and molecules. Exquisite control over the dense internal structure of molecules, in particular, offers promising avenues for tests of fundamental physics \cite{Andreev2018,Lim2018,Kozyryev2017,Kobayashi2019,Carollo2019,Borkowski2019}, cold controlled chemistry and collisions \cite{Krems2008,Balakrishnan2016,Bohn2017,NiHuScience19_KRbReactions,Guo2018,McDonald2016}, quantum simulation \cite{Blackmore2018,Altman2019,Micheli2006}, quantum computation \cite{Ni2018,Hudson2018,Hughes2019,Yu2019}, and compact THz frequency references \cite{Wang2018}.
Addressing the internal states with high fidelity necessitates the manipulation of spatial and motional degrees of freedom. For cold neutral gases, one common approach is the use of optical traps. Given two molecular states, the differential trap-induced light shifts and motional decoherence can be eliminated by tuning the frequency of the trap light close to a narrow rovibronic transition such that the dynamic polarizabilities are equal. Such magic wavelength traps were recently demonstrated to extend the vibrational \cite{Kondov2019} and rotational \cite{Bause2019,Seeselberg2018} coherence times of an ensemble of ultracold molecules. As with atomic lattice clocks \cite{atomic1,atomic2}, scattering of lattice laser light is expected to play a central role in limiting achievable interrogation times. Accurate knowledge of the molecular structure is thus a primary step in the identification of specific loss channels involved in the interaction of the molecules and the trap light.  Additionally, this knowledge will help determine feasible pathways for quantum state preparation of the molecules \cite{park2015two, Guo2017, Aikawa2011, ciameipra2017}.

In this work, we employ Lamb-Dicke spectroscopy in an optical lattice and state-of-the-art \textit{ab initio} calculations to measure and predict vibronic line strengths in $^{88}\mathrm{Sr}_2$ molecules spanning over three orders of magnitude. This is achieved by measuring light shifts induced by a coupling laser as it is swept across a transition of interest, resulting in avoided crossing curves from which the line strengths are extracted with minimal modeling. We probe transitions between the electronic ground potential X$^1\Sigma_g^+$ (henceforth referred to as X) and singly-excited Hund's case (c) potentials $(1) 0_u^+$ and $(1) 1_u$ corresponding to the $^1S_0$-$^3P_1$ dissociation limit. By addressing both weakly and deeply bound states, the molecular potential energy curves are probed over a wide range of internuclear distances. We combine our measurements with spectroscopic data to construct refined potential curves for $0_u^+$ and $1_u$ in the Morse/Long-range form, and apply the results to judiciously select magic wavelengths that alleviate the impact of both frequency instability and scattering of the lattice laser on coherent molecule-light interactions.  We demonstrate a superposition of far-separated vibrational states that remains coherent for a record time of nearly 100 ms.

\begin{figure}
    \centering
    \includegraphics[width=\columnwidth]{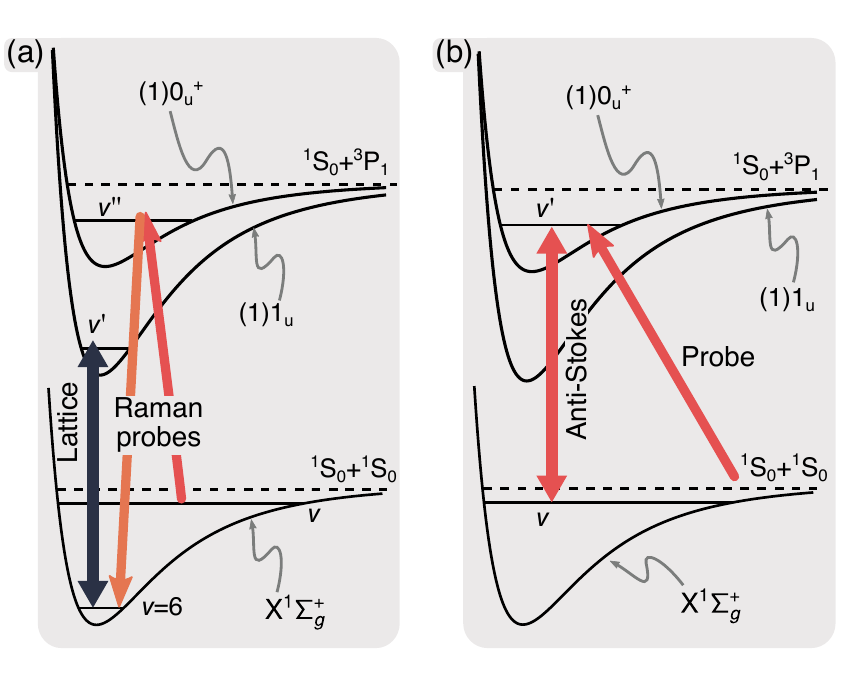}
    \caption{Line strength measurement schemes for (a) deeply and (b) weakly bound states ($v,v'$).  Spectroscopic probing is performed on lattice-trapped ultracold X$^1\Sigma_g^+$ molecules in the $v=-1$ or $-2$ vibrational states in (a), and on $^1S_0$ atoms in (b).}
    \label{fig:combined}
\end{figure}

In the Born-Oppenheimer approximation, the wave function of a diatomic molecule is a product of its electronic and rovibrational parts. The rovibrational part $|\chi^{v, J}_{n, \Omega}(R) \rangle$ is a function of the internuclear distance $R$ and is labeled by the following quantum numbers: the electronic channel $n$, the vibrational number $v$, the total angular momentum $J$, and its projection onto the internuclear axis in the molecule-fixed frame $\Omega$. Following Ref.~\cite{Hansson2005}, the line strength (or transition strength) for an electric-dipole transition between the rovibrational states described by the wave functions  $|\chi^{v, J}_{n, \Omega}(R) \rangle$ and $|\chi^{v', J'}_{n', \Omega'}(R) \rangle$ is given by,
\begin{align}
\label{eq:Sdef}
 S \equiv |H_{J' M' \Omega'}^{J M \Omega} \langle \chi^{v', J'}_{n', \Omega'}(R) | d_{\Omega'-\Omega}(R) | \chi^{v, J}_{n, \Omega}(R) \rangle|^2,
\end{align}
where $d_{\Omega'-\Omega}(R)$ is the electronic transition dipole moment, $M$ is the projection of the total angular momentum onto the lab-frame $Z$ axis, and $ H^{J M \Omega}_{ J'M' \Omega'}$ is the rotational factor defined in the Supplemental Material \cite{thispapersuppmat}. In the case of transitions driven by linearly polarized light along the quantization axis, the selection rules force $M' = M$. Eq.~(\ref{eq:Sdef}) can easily be generalized to the multichannel case by introducing a sum over $n$, $n'$, $\Omega$ and $\Omega'$ before taking the absolute square. The quantity of interest that is readily accessible is the Rabi frequency
\begin{equation}
f_R = \frac{1}{h}\sqrt{\frac{2IS}{\epsilon_0 c}}, \end{equation}
which quantifies the strength of coupling between two states when driven by a laser with irradiance $I$, leading to observable light shifts in the molecular spectra. In this study, we measure the light shifts induced either by the optical lattice which also acts as the trap as in Fig.~\ref{fig:combined}(a), or by an anti-Stokes laser as in Fig.~\ref{fig:combined}(b).
    
\begin{figure}
    \centering
    \includegraphics[width=\columnwidth]{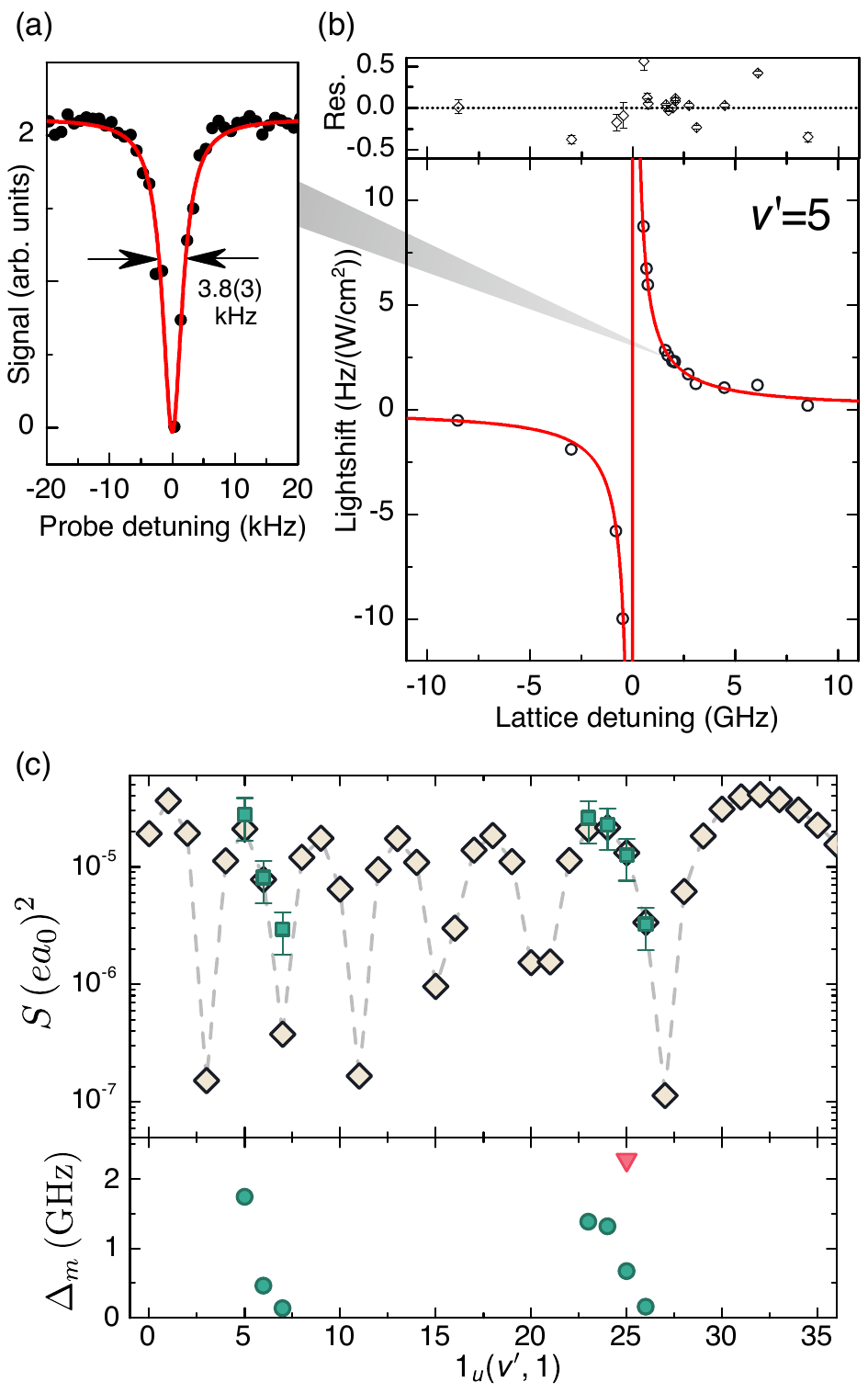}
    \caption{(a) A two-photon Raman depletion lineshape. For this trace, the differential light shift of the X states is nearly nulled, resulting in a narrow power-broadened linewidth of 3.8(3) kHz inferred from a Lorentzian fit. (b) The Raman peak locations (open circles) exhibit dispersive behavior as the lattice frequency is swept across the $1_u$ state (shown here for $v^\prime = 5$). The solid red line is the fit to the avoided crossing of the form $f_R^2/4\Delta$. Residuals have units of Hz/(W/cm$^2$). Error bars denote 1$\sigma$ uncertainties. (c)  Line strengths of $\mathrm{X}(6,0)\rightarrow 1_u(v^\prime,1)$ for various $v^\prime$ near the potential minimum (green squares:  measured; yellow diamonds: calculated using the Morse potential). The errors bars here include the systematic uncertainty of the beam waist. Also shown are the magic lattice detunings relative to the nearest $1_u$ state for a given pair of $\mathrm{X}$ states (green circles: $\mathrm{X}(-1,0)\rightarrow \mathrm{X}(6,0)$; red triangle: $\mathrm{X}(-1,0)\rightarrow \mathrm{X}(4,0)$), as listed in Table \ref{tab:magicdet} \cite{thispapersuppmat}}
    \label{fig:deep}
\end{figure}

In the scheme shown in Fig.~\ref{fig:combined}(a), we explore the coupling of X$(v=6,J=0)$ to a set of $J^\prime=1$ resonances near the bottom of the $1_u$ potential well by the optical lattice. For a given lattice detuning $\Delta$ from the $1_u$ state and the corresponding Rabi frequency $f_R$, the additional light shift on X$(6,0)$ is $f_R^2/4\Delta$ in the limit $\Delta \gg f_R$. To probe this shift, we perform two-photon Raman spectroscopy on X$(6,0)$, shown in Fig.~\ref{fig:deep}(a). We initialize our molecules in either X$(-1,0)$ or X$(-2,0)$ via photoassociation \cite{thispapersuppmat}, and use $0_u^+(v^{\prime\prime}=-4,J^{\prime\prime}=1)$ or $0_u^+(-5,1)$ as the intermediate state respectively. The Raman lasers are detuned from the intermediate state by $>$20 MHz, and the frequency of the first Raman leg is swept while the second is kept fixed. The corresponding light shift on X$(-1,0)$ or X$(-2,0)$ is less than $30$ mHz, thus the choice of initial state does not strongly affect the measured $S$. Measurement of the two-photon resonance frequency at various $\Delta$ gives the expected dispersive behavior, as shown in Fig.~\ref{fig:deep}(b). 

Using this technique, we measure seven line strengths from $\mathrm{X}(6,0)$ to $1_u(v^\prime,1)$, where all the $1_u$ states are below the minimum of the $0_u^+$ potential. As illustrated in Fig. \ref{fig:deep}(c), experimentally the line strengths $S$ demonstrate decreasing trends in two ranges. Prior to this work, we used the binding energies calculated using the \textit{ab initio} model of Ref.~\cite{Skomorowski2012jcp} to assign the vibrational quantum numbers $v^\prime$ in the ranges 1--3 and 19--22, where we find good agreement within 0.5\% of the experimental binding energies (see Table \ref{tab:bindingenergies} \cite{thispapersuppmat}). To calculate the line strengths, we model X with a set of precise empirical potential parameters obtained from the hot pipe Fourier-transform spectroscopy \cite{Stein2010} and use the \textit{ab initio} electronic transition dipole moment. To our surprise, the calculated \textit{ab initio}
line strengths follow the opposite trends compared to the observations, and this persisted even when the electronic transition dipole moment was modified. This suggests that the $v^\prime$ assignment based on the \textit{ab initio} $1_u$ potential is erroneous. To overcome this, we take an alternative approach and model the short-range behavior of $1_u$ with the simple Morse potential \begin{equation}\label{eq:simpleMorse}
    V(R) = D_e \left[1-\mathrm{e}^{-\beta (R-R_e)}\right]^2 -D_e,
\end{equation}
where $D_e$ is the potential depth, $R_e$ is the equilibrium bond length, $\beta \equiv 2\pi\sqrt{2\mu \omega_e x_e c/h}$, and $\mu$ is the reduced mass of the dimer. In the presence of vibrational-rotational interaction, the energy levels are \begin{align}\label{eq:vibrotor} E(v^\prime,J^\prime) = -D_e + \omega_e\left(v^\prime+\frac{1}{2}\right) - \omega_e x_e\left(v^\prime+\frac{1}{2}\right)^2 \nonumber\\
+ \left[B_e-\alpha_e \left(v^\prime+\frac{1}{2}\right)\right]J^\prime(J^\prime+1),\end{align} where $\omega_e$, $x_e$, $B_e$ and $\alpha_e$ are the vibrational, anharmonicity, rotational and vibration-rotation coupling spectroscopic constants respectively. As shown in the Supplemental Material \cite{thispapersuppmat}, the best fit of the seven observed $J^\prime = 1$ levels to Eq.~(\ref{eq:vibrotor}) indicates 15 vibrational states between the two groups and $\omega_e x_e = 0.2123(17)\,\mathrm{cm}^{-1}$. Rotational spectroscopy of $J^\prime = 3$ states yields $\alpha_e = 7.0652(59) \times 10^{-5} \,\rm{cm}^{-1}$. The Morse eigenfunctions were obtained by numerically solving the nuclear Schr\"{o}dinger equation on an adaptive grid \cite{nucleargrid}. As the classical turning points of the deeply bound X states are much further apart than those of $1_u$, the Frank-Condon factors between them largely depend on the spatial variation of the ground state nuclear wavefunction. Consequently, the calculated $S$ versus $v^\prime$ exhibit a characteristic interference-like pattern with $v+1$ maxima. The observed experimental trends in $S$ are well captured by the calculations only for $v^\prime$ assignments of 5--7 and 23--26 for the two ranges, as shown in Fig.~\ref{fig:deep}(c). The corresponding potential parameters are $D_e  = 6387.89(11) \,\mathrm{cm}^{-1}$ and $R_e = 7.9027(5)\, a_0$. 

Hence, while the \textit{ab initio} calculation has good accuracy in the long range, it underestimates $D_e$ by approximately 300 cm$^{-1}$ (5\% relative difference). On the other hand, Eq.~(\ref{eq:simpleMorse}) allows for $D_e$ to be empirically determined but the simple potential cannot be extrapolated to the long range. To combine these complementary descriptions, we recast the \textit{ab initio} $1_u$ and $0_u^+$ potentials in the Morse/Long-range (MLR) form \cite{LeRoy2009,Roy2007} and fit to spectroscopic data, while fixing $D_e$ and $R_e$ to their empirical values found in this work. In the Supplemental Material \cite{thispapersuppmat}, we provide details of the fitting process and the MLR parameters, and recalculate the energy levels of $0_u^+$ and $1_u$ to benchmark against experimental values.

\begin{figure}
    \centering
    \includegraphics[width=\columnwidth]{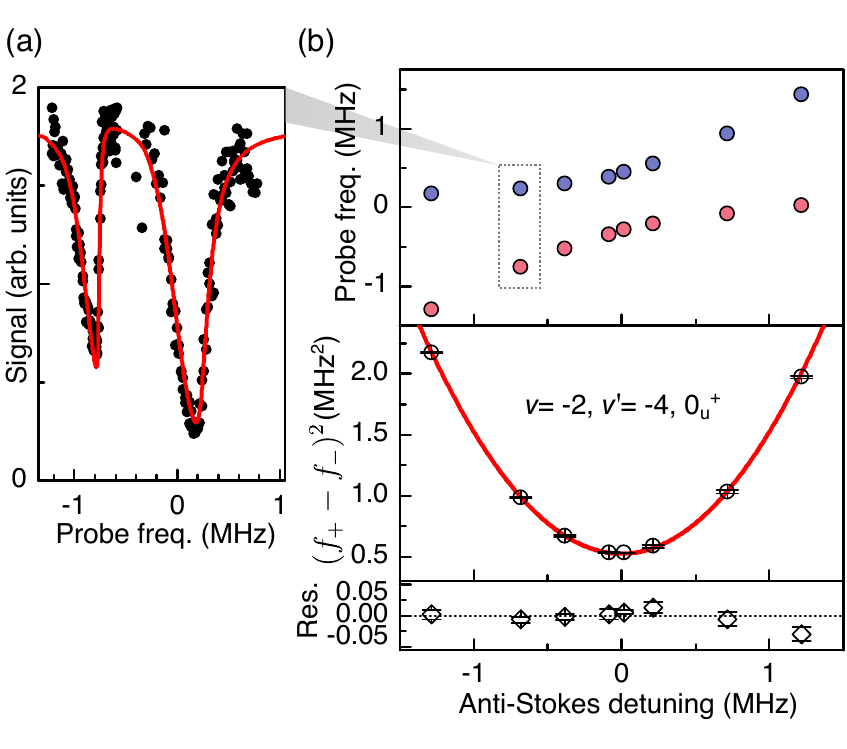}
    \caption{(a) For a given detuning of the anti-Stokes laser from the bound-to-bound molecular transition, the probe is scanned to obtain an Autler-Townes doublet. (b) The locations of the blue-side peak (blue circles) and the red-side peak (red circles) form an avoided crossing. The square of the peak separations fit to a parabola (solid red) when plotted against the anti-Stokes detuning, the minimum of which is $f_R^2$. Error bars are propagated from 1$\sigma$ uncertainties in the peak locations from the lineshape fit. Residuals are in units of MHz$^2$.}
    \label{fig:shallow}
\end{figure}

For weakly bound states, an anti-Stokes laser selectively dresses X$(v, J=0)$ with a ro-vibrational state $0_u^+(v^\prime, J^\prime=1)$ or $1_u(v^\prime, J^\prime=1)$, as shown in Fig.~\ref{fig:combined}(b). Here, we perform spectroscopy on a trapped sample of ultracold \textit{atoms}. Depletion occurs when a weak probe laser is on resonance with the dressed states \begin{equation}\label{eq:fpm}
    f_\pm = \frac{\Delta}{2}\pm\frac{\sqrt{f_R^2+\Delta^2}}{2},
\end{equation} where $f_+ (f_-)$ is the resonance frequency of the blue-side (red-side) peak relative to the bare resonance ($f_R = 0$), and $\Delta$ is the coupling laser detuning. Fig.~\ref{fig:shallow}(a) shows a sample trace when the anti-Stokes laser is slightly red-detuned from the X$(-2, 0)\rightarrow 0_u^+(-4, 1)$ transition, revealing an Autler-Townes doublet. Although the atoms are tightly confined along the axial direction of the 1D lattice, the radial confinement is much weaker. At a finite temperature of a few microkelvin, the atoms occupy several motional states above the $^1S_0$+$^1S_0$ continuum leading to an asymmetric lineshape. We determined the location of the resonances by fitting to a doublet lineshape function that accounts for these thermal effects \cite{McGuyer2015}.

Keeping the anti-Stokes laser intensity constant, the square of the doublet separation $(f_+-f_-)^2$ versus $\Delta$ is a parabola whose minimum is $f_R^2$, as shown in Fig. \ref{fig:shallow}(b). This presents an attractive method of determining line strengths, as opposed to measurements involving power broadening and transition rates, since frequency differences are robust against a wide variety of effects such as reference cavity drift, the overall trap-induced light shift, shot-to-shot signal fluctuations, and the fit lineshape for the doublet. Moreover, by working strictly with transitions between $J=0$ and $J^\prime = 1$ and in the regime where the anti-Stokes detunings are larger than the Zeeman structure, the measured $S$ are insensitive to laser polarization and are effectively between $M=M^\prime=0$ states. The measured $S$ and corresponding predictions from the \textit{ab initio} \cite{Skomorowski2012jcp} and MLR models are listed in Table \ref{tab:0ushallow}. For the weakly Coriolis-mixed states both models perform similarly well. However, for the strongly Coriolis-mixed states, only the MLR model gives the correct $0_u^+$ or $1_u$ assignments and thus is more accurate in its predictions for $S$.

\begin{table}
 \caption{Measured line strengths for weakly bound $0_u^+$ and $1_u$ states from X for various vibrational pairs. Also shown for comparison are predictions from the \textit{ab initio} (AI) and adjusted Morse/Long-range (MLR) potential constructed in this work. Starred values are strongly Coriolis-mixed states. Statistical uncertainties account for the 1$\sigma$ errors in the extracted Rabi frequencies as well as for laser power fluctuations.  The units are $10^{-3} (ea_0)^2$.}
    \label{tab:0ushallow}
\begin{ruledtabular}
    \centering
    \begin{tabular}{llllll}
    X$(v,J=0)$&State&$(v^\prime,J^\prime=1)$&AI&MLR&Exp.\\
    \colrule
    -1&$0_u^+$&-4&3.09&2.77&2.57(4)\\
    -2&$0_u^+$&-4&0.81&0.74&0.70(2)\\
    -2&$0_u^+$&-5&5.86&5.06&4.30(6)\\
    -3&$0_u^+$&-6&0.07$^*$&8.89&8.7(4)\\
    -1&$1_u$&-1&5.44&4.56&5.53(8)\\
    -1&$1_u$&-2&0.36&0.33&0.40(1)\\
    -2&$1_u$&-1&1.71&1.68&1.74(3)\\
    -2&$1_u$&-2&6.95&5.82&8.0(1)\\
    -3&$1_u$&-3&13.2$^*$&2.46&2.10(5)\\
    \end{tabular}
    \end{ruledtabular}
\end{table}

 \begin{figure}
    \centering
    \includegraphics[width=8.6cm]{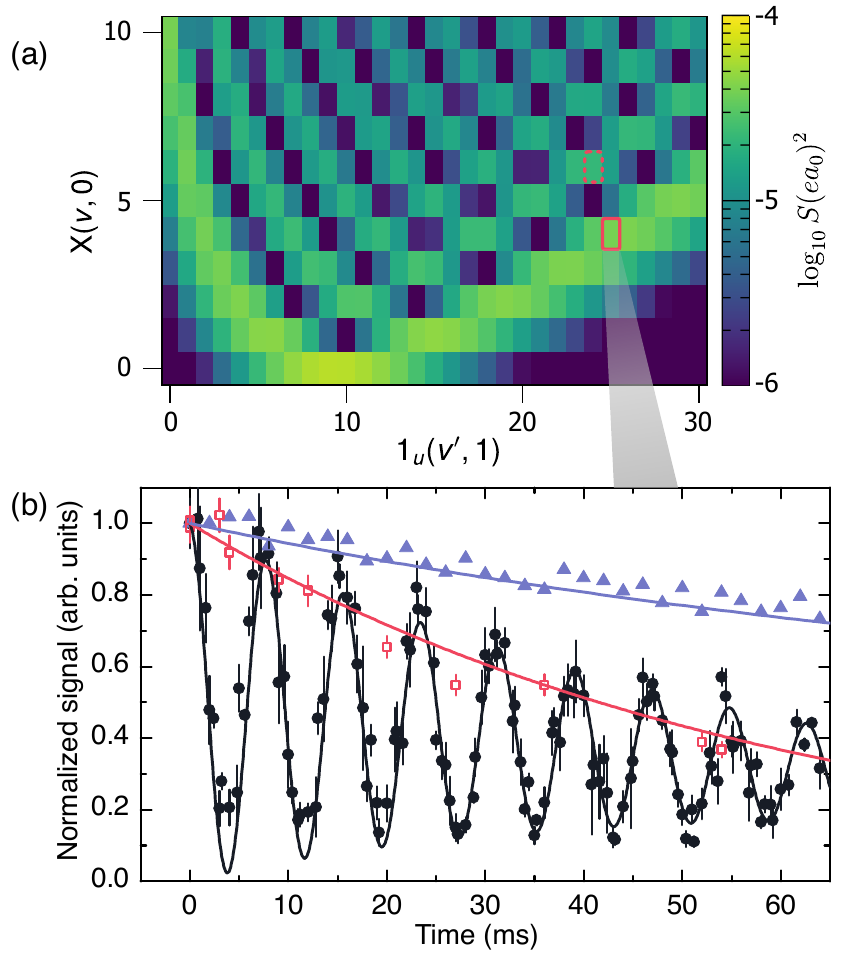}
    \caption{(a) Line strengths of deeply bound states of $1_u$ to X. Solid rectangle: the transition used for the magic wavelength in this work. Dashed rectangle: our previous work \cite{Kondov2019}. (b) Two-photon Rabi oscillations between X$(-1,0)$ and X$(4,0)$ (black circles). Here, favorable magic trapping is achieved by tuning the lattice near the X$(4,0) \rightarrow 1_u(25,1)$ transition. Also shown are the normalized population decay of X$(4,0)$ (red squares) and X$(-1,0)$ (blue triangles). Black line: fit to $A\exp(-t/T_1)\left[1+\exp(-t/T_2^{Rabi})\cos(\omega t -\phi)\right]$. Red and blue lines are fits to the rate equation $\dot{N} = -k_\gamma N^\gamma$ for the molecular number $N$ with $\gamma =$ 1 and 2 respectively and $k_\gamma$ as a free parameter. Error bars are 1$\sigma$ uncertainties.}
    \label{fig:dms1u}
\end{figure}
 
Our findings directly inform the engineering of favorable magic-wavelength optical traps for a molecular clock, by means of elucidating the quantum chemistry of the strontium dimer. Just as in atomic lattice clocks, for a given baseline polarizability mismatch between the clock states, the required magic detuning $\Delta_m$ (relative to a resonance between one of the clock states and an excited state) monotonically increases with the line strength (see Fig.~\ref{fig:deep}(c) and Tab.~\ref{tab:magicdet} \cite{thispapersuppmat}). The sensitivity of the clock transition to lattice frequency inaccuracies is simply the slope of the lattice-induced light shift at the magic detuning, $-f_R^2/4\Delta_m^2$, and would decrease monotonically for larger $S$ (and hence larger $\Delta_m$). Therefore, magic wavelengths based on stronger transitions place less stringent constraints on the required frequency stability of the lattice laser and on the bandwidth of the spectral filter that suppresses the lattice laser noise away from the carrier (such as amplified spontaneous emission). To this end, we compute $S$ between the lower-lying $J^\prime=1$ states of $1_u$ and several deeply bound $J=0$ states of X, as shown in Fig.~\ref{fig:dms1u}(a). For X$(4,0)$, the state $1_u(25,1)$ has one of the largest predicted line strengths among several $v^\prime$ in the vicinity. The measured magic frequency at 330.302,670,449(104) THz corresponds to a detuning $\Delta_m = 2.277(18) \,\mathrm{GHz}$ from the X$(4,0) \rightarrow 1_u(25,1)$ resonance, and is the largest studied in this work. Operating the molecular clock at this magic wavelength, we demonstrate long-lived two-photon Rabi oscillations ($T_2^{Rabi}=77(6)$ ms, $T_1=127(8)$ ms) between the clock states X$(-1,0)$ and X$(4,0)$, as shown in Fig.~\ref{fig:dms1u}(b) and described in the caption. This represents a significant improvement in coherent light-molecule interactions over our previous experiment \cite{Kondov2019}. The oscillations are predominately damped by the loss of X$(4,0)$ molecules, which has a $1/e$ lifetime of 60(2) ms at a trap depth of $k_B \times 12(1) \,\mu\mathrm{K}$. The single-body losses are faster for deeper traps, which indicates scattering of the lattice light. Here, care must be taken to account for the linewidth of the lattice trap laser, since the deeply bound $1_u$ states are expected to be much narrower ($\Gamma_e \lesssim 2\pi \times 6 \,\mathrm{kHz}$), and the situation is that of broadband scattering. Further investigations using the MLR potential curves can help lend credence to specific loss mechanisms such as multi-photon scattering or to rule them out quantitatively. 

In summary, we have measured the line strengths of several $0_u^+$ and $1_u$ states connecting to the ground-state X potential in two different regimes, $-$ weakly bound near-threshold states and deeply bound states, $-$ of ultracold lattice-trapped Sr$_2$ molecules used in a molecular clock. The measurements were used to obtain analytic MLR potential curves for $0_u^+$ and $1_u$ that are in good agreement with previously published binding energies and those found in this work. In particular, we demonstrate the reliability of the constructed $1_u$ potential by predicting and verifying states that have large transition strengths to X, which is an important criterion for the construction of magic optical traps. We presented an improved choice of a magic trap that led to a coherent control of the clock states for a record duration of nearly 100 ms.
Furthermore, our accurate model will help elucidate the processes that contribute to the quenched molecular lifetimes, and inform stimulated adiabatic pathways for ground-state preparation in future work, raising the prospects for a comprehensive vibrational spectroscopy of the ground potential as a high-precision test of molecular quantum electrodynamics and possible new physics. 

\begin{acknowledgments}
We are grateful to A. Liberman and Y. Chai for contributions to the experiment. We acknowledge support from NSF grant No. PHY-1911959 and ONR grant No. N00014-17-1-2246.  R. M. and I. M. also acknowledge the Polish National Science Center Grant No. 2016/20/W/ST4/00314.
\end{acknowledgments}

%


\clearpage
\pagebreak

\appendix

\setcounter{equation}{0}
\setcounter{figure}{0}
\setcounter{table}{0}
\setcounter{page}{1}
\makeatletter
\renewcommand{\theequation}{S\arabic{equation}}
\renewcommand{\thefigure}{S\arabic{figure}}
\renewcommand{\thetable}{S\Roman{table}}
\section{SUPPLEMENTAL MATERIAL}
\section{Experimental setup}

A detailed account of our experimental setup has been given elsewhere \cite{Reinaudi2012}. Briefly, $^{88} \mathrm{Sr}$ atoms are laser-cooled in a two-stage magneto-optical trap operating on the strong $^1S_0$-$^1P_1$ transition at 461 nm and the spin-forbidden intercombination transition $^1S_0$-$^3P_1$ at 689 nm, producing over $10^4$ atoms in the $^1S_0$ electronic ground state at a final temperature of approximately $4(1)\,\mu \mathrm{K}$ as measured from a ballistic expansion of the gas. The atoms are transferred to a horizontal one-dimensional optical lattice with a typical trap depth of $50\,\mu \mathrm{K}$ and axial confinement frequency of 70 kHz such that spectroscopy near the intercombination line is in the Lamb-Dicke and resolved sideband regimes. Note that for the data demonstrating Rabi oscillations, we used a lower trap depth of $12\,\mu \mathrm{K}$ to prolong the lifetime of the deeply bound X state.

We can efficiently produce weakly bound vibrational states belonging to the ground potential, X, by photoassociating to a weakly bound $0_u^+$ state where a reasonably strong spontaneous decay pathway to X exists owing to favorable Franck-Condon overlap. Depending on the molecular transition of interest, we perform spectroscopy on a sample of either ultracold strontium atoms or molecules. We count the number of atoms remaining in $^1S_0$ after a spectroscopy sequence by absorption imaging on the $^1S_0$-$^1P_1$ transition (a molecular sample is first photodissociated into atoms). 

The photoassociation, photodissociation, and probe laser beams are phase-locked to a sub-kHz linewidth master laser that is itself stabilized to a high-finesse reference cavity ($\mathcal{F} \sim$ 30,000). The Raman lasers are co-stabilized with an erbium-doped fiber-based optical frequency comb using the transfer-oscillator technique \cite{Stenger2002,Scharnhorst2015}. The repetition rate of the frequency comb is phase-locked to the master laser, while its carrier offset is referenced to a GPS guided Rb frequency standard. For the coherent Rabi oscillations, the lattice is phase-locked to the frequency comb. All laser beams are coaligned with the lattice and focused onto the atoms or molecules. 

\section{Optical beam modeling \label{sec:beam}}
To determine the spatial profile of the light shift inducing laser beam (either the anti-Stokes or the optical lattice), we deflect the beams just before the viewport of the vacuum chamber onto a camera (Thorlabs DCC1545M) positioned at the focal plane of the forward-pass lattice beam. Typically, the waist of the lattice beam is $<50\,\mu\mathrm{m}$, and that of the anti-Stokes is much larger at $100\,\mu\mathrm{m}$. The pixel size of the camera is $5.2\,\mu\mathrm{m}\equiv h$. 

In the measurements involving weakly bound states, to account for any possible misalignment of the anti-Stokes and any non-idealities in its spatial profile, the local intensity experienced by the molecules is estimated as the intensity in one square pixel at the position of the lattice on the camera. This was achieved by obtaining a conversion ratio between the total count read by all camera pixels ($C_{\mathrm{tot}}$) to the total optical power received on a power meter ($P$) in the same optical path. The local intensity is thus \begin{equation}
    \frac{P}{C_{\mathrm{tot}}}\frac{C_{\mathrm{local}}}{h^2},
\end{equation} where $C_{\mathrm{local}}$ is the pixel count at the lattice position. For an ideal Gaussian beam with waist $w_0$, this method systematically underestimates the intensity near the peak due to the finite pixel size by a factor \begin{equation}
    \frac{\pi}{2}\left[\frac{w_0}{h}\mathrm{erf}\left(\frac{h}{\sqrt{2}w_0}\right)\right]^2.
\end{equation} For our parameters, this error propagates into the line strength at the $0.01\%$ level and is negligible compared to the statistical error arising from the measurement of the Rabi frequency and the optical power.

In the measurements involving deeply bound states, the lattice trap laser itself induces the light shift. Here, we take the intensity experienced by the molecules to be the peak intensity of the lattice, which is determined by the waists and alignment of the counter-propagating beams. We fit a Gaussian spatial profile to the forward-pass lattice beam to extract its beam waist, $w_\mathrm{latt}$. This is converted to an effective waist, $w_\mathrm{eff}$, which takes into account fluctuations in the quality of the beam alignment due to, for example, mechanical creep of the retro-reflecting mirror, or imperfect wavefront matching of the retroreflected beam with the forward beam. From lattice sideband spectroscopy, we know that $w_\mathrm{latt} \lesssim w_\mathrm{eff} \lesssim \sqrt{2} w_\mathrm{latt}$, and thus take the the average and range as respectively the value and uncertainty for $w_\mathrm{eff}$. The local intensity, in this case, is \begin{equation}
    \frac{8P_f}{\pi w_\mathrm{eff}^2}, 
\end{equation} where $P_f$ is the optical power of the forward pass beam. For the results reported here, $w_{\mathrm{eff}} = 42(7)$ $\mu$m. This is the dominant source of error in our measurements of $S$ for the deeply bound $1_u$ states. 

\section{Definition of the Rotational Factor}

The rotational factor for the line strength is defined as
 
\begin{align}
 H^{J M \Omega}_{ J'M' \Omega'}\equiv & (-1)^{M-\Omega}
 \sqrt{(2J+1)(2 J'+1)} \\ \nonumber & \times\sqrt{1+\delta_{\Omega0}+\delta_{\Omega'0} -2  \delta_{\Omega0} \delta_{\Omega'0}}  
 \\ \nonumber & \times 
   \begin{pmatrix}
  J & 1 & J'\\
  M &0 & -M'
 \end{pmatrix}
    \begin{pmatrix}
  J & 1 & J'\\
  \Omega &\Omega'-\Omega & \Omega'
 \end{pmatrix}. 
\end{align}

\section{Measured magic detunings and line strengths}

The differential polarizability between the X ``clock" states is nulled when the lattice laser frequency is such that it is detuned with respect to a $\mathrm{X}\rightarrow 1_u$ transition by a magic detuning $\Delta_m$. One clock state will thus have its polarizability strongly modified, while leaving the second clock state mostly unperturbed. In this study, we always choose to tune the polarizability of the more deeply bound vibrational clock state, i.e. either $\mathrm{X}(6,0)$ or $\mathrm{X}(4,0)$.  In Tab. \ref{tab:magicdet} we list the experimentally measured detunings $\Delta_m$ relative to various $\mathrm{X}\rightarrow 1_u$ transitions, along with the corresponding line strengths. As expected, the magic detunings increase monotonically for increasingly stronger $\mathrm{X}\rightarrow 1_u$ transitions. 

\begin{table}
\caption{List of measured magic detunings and $\mathrm{X}\rightarrow 1_u$ line strengths.}
 \label{tab:magicdet}
 \begin{ruledtabular}
    \centering
    \begin{tabular}{cccc}
Clock states& $\mathrm{X}\rightarrow 1_u$& $\Delta_m$ (GHz)& $S\, (10^{-5}\,(ea_0)^2)$\\
\colrule
$\mathrm{X}(-1,0) \rightarrow \mathrm{X}(6,0)$&$1_u(26,1)$&0.160(4)& 0.32(13)\\
$\mathrm{X}(-1,0) \rightarrow \mathrm{X}(6,0)$&$1_u(25,1)$&0.670(18)&1.25(49)\\
$\mathrm{X}(-1,0) \rightarrow \mathrm{X}(6,0)$&$1_u(24,1)$&1.3157(12)&2.27(89)\\
$\mathrm{X}(-1,0) \rightarrow \mathrm{X}(6,0)$&$1_u(23,1)$&1.380(2)&2.6(1.0)\\
$\mathrm{X}(-1,0) \rightarrow \mathrm{X}(6,0)$&$1_u(7,1)$&0.135(14)&0.29(12)\\
$\mathrm{X}(-1,0) \rightarrow \mathrm{X}(6,0)$&$1_u(6,1)$&0.46(10)&0.81(32)\\
$\mathrm{X}(-1,0) \rightarrow \mathrm{X}(6,0)$&$1_u(5,1)$&1.743(13)&2.8(1.1)\\
\colrule
$\mathrm{X}(-1,0) \rightarrow \mathrm{X}(4,0)$&$1_u(25,1)$&2.277(18)&9.4(3.2)
    \end{tabular}
    \end{ruledtabular}
\end{table}

\section{Spectroscopic constants of the $1_u$ potential \label{sec:spectroscopic}}

The energy spectrum of the simple Morse potential with vibration-rotation coupling is \begin{align}\label{eq:vibrotorsupp} E(v^\prime,J^\prime) = &-D_e + \omega_e\left(v^\prime+\frac{1}{2}\right) - \omega_e x_e\left(v^\prime+\frac{1}{2}\right)^2 \nonumber\\
&+ \left[B_e-\alpha_e \left(v^\prime+\frac{1}{2}\right)\right]J^\prime(J^\prime+1),\end{align} where $\omega_e$, $x_e$, $B_e$ and $\alpha_e$ are the vibrational, anharmonicity, rotational and vibration-rotation coupling spectroscopic constants respectively.

\begin{figure}
    \centering
    \includegraphics[width=\columnwidth]{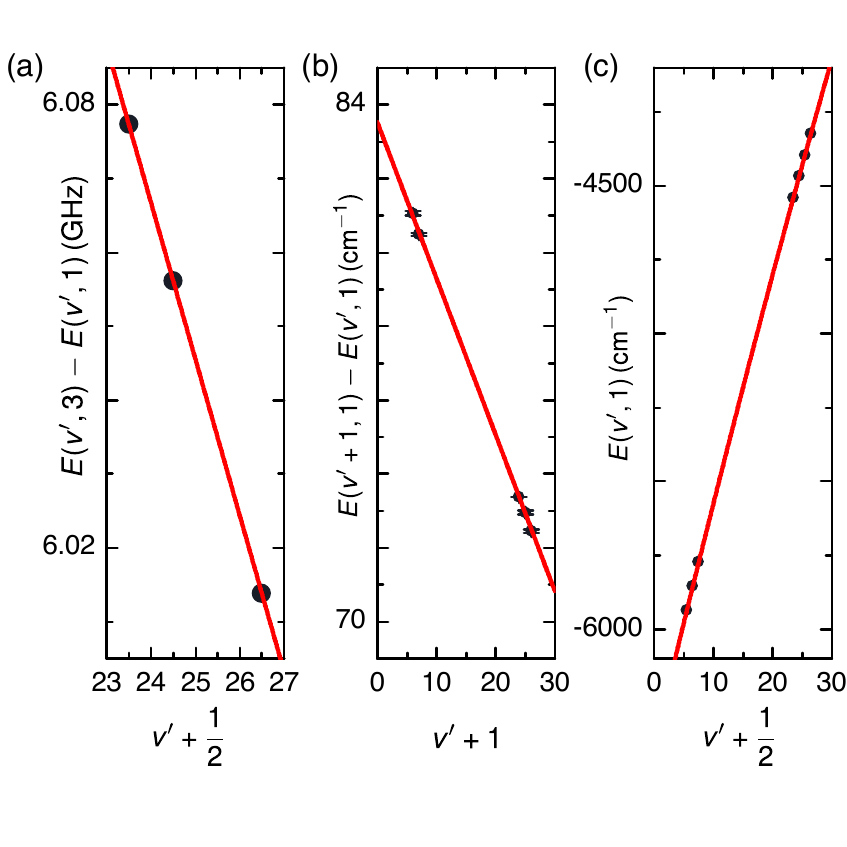}
    \caption{(a) Rotational splittings between $J^\prime =3$ and 1.  (b) Adjacent vibrational splittings of $J^\prime=1$. (c) Binding energies of $J^\prime=1$. All states belong to the $1_u$ potential. The red lines are fits to the data: a linear function in (a) and (b), and a quadratic function in (c). The intercepts with the vertical axis are $10B_e$, $(\omega_e - 2\alpha_e)$, and $2B_e-D_e$ respectively. Error bars are $1\sigma$ uncertainties  in (b), and are much smaller than the symbol size in (a) and (c). }
    \label{fig:specparams}
\end{figure}

To find $\alpha_e$, we measure the rotational splitting of $J^\prime =3$ and 1 for various $v^\prime$. Using Eq.~(\ref{eq:vibrotorsupp}), we see that \begin{equation}
    E(v^\prime,3) - E(v^\prime,1) = 10B_e - 10\alpha_e\left(v^\prime+\frac{1}{2}\right),
\end{equation} so that plotted against $v^\prime+\frac{1}{2}$ the slope of the linear fit is $-10\alpha_e$ and insensitive to an overall offset in the $v^\prime$ labels. In order to address both $J^\prime=3$ and 1, we populate X$(v=6,J=2)$ with a long stimulated Raman pulse from X$(v=-1,J=0)$ in a non-magic lattice.  Then, a short probe pulse ($\sim 100\,\mu\rm{s}$) resonant with either $J^\prime$ state depletes the population. Finally, we perform another Raman pulse to transfer the population back to X$(-1,0)$ where it is dissociated and detected. The Raman transfer is inefficient owing to the polarizability mismatch, but we nevertheless obtain sufficient signal to perform the spectroscopy. We vary the optical lattice power between two extremes and linearly extrapolate to zero power to obtain the resonance frequency. The absolute frequency of a resonant probe pulse was determined using an optical frequency comb. Tab.~\ref{tab:rotsplit} lists the measured rotational splittings for three states, and Fig.~\ref{fig:specparams}(a) shows the linear fit from which we extract $\alpha_e = 7.068(11) \times 10^{-5} \,\rm{cm}^{-1}$. 

\begin{table}
\caption{Rotational splitting (in units of GHz) between $J^\prime=3$ and $J^\prime=1$ for a given $v^\prime$ state of $1_u$.}
 \label{tab:rotsplit}
 \begin{ruledtabular}
    \centering
    \begin{tabular}{cc}
$v^\prime$&$E(v^\prime,3) - E(v^\prime,1)$\\
\colrule
23&6.07728(9)\\
24&6.05616(9)\\
26&6.01375(6)
    \end{tabular}
    \end{ruledtabular}
\end{table}

To determine $\omega_ex_e$, we take differences of adjacent vibrational energy levels for $J^\prime =1$. Eq.~(\ref{eq:vibrotorsupp}) suggests that \begin{equation}
    E(v^\prime+1,1) - E(v^\prime,1) = (\omega_e - 2\alpha_e) - 2\omega_ex_e(v^\prime +1).
\end{equation} Therefore, plotting this difference against $v^\prime+1$ should yield a straight line with a slope equal to $-2\omega_ex_e$ that is similarly insensitive to an overall offset in the $v^\prime$ labels (Fig.~\ref{fig:specparams}(b)). The best fits, judged on the basis of the reduced $\chi^2$, are obtained if there are 15 intermediate vibrational states in the region where data is absent. We thus obtain $\omega_e x_e = 0.21150(28)\,\mathrm{cm}^{-1}$.

Note that for the linear fits in Fig.~\ref{fig:specparams} (a) and (b), the vertical intercepts are $10B_e$ and $(\omega_e - 2\alpha_e)$, respectively, \emph{only if} accurate knowledge of the $v^\prime$ labels are available. In this work, we rely on finding potential parameters that reproduce the trends in the line strengths to unambiguously assign the levels as $v^\prime$ = 5--7, 23--26, as described in the main text. With this at hand, we extract $B_e = 0.021933(3)\,\mathrm{cm}^{-1}$ and $\omega_e = 83.528(13) \,\mathrm{cm}^{-1}$. The equilibrium bond length is then calculated using $R_e = \frac{1}{2\pi} \sqrt{\frac{h}{2\mu cB_e}} = 7.9027(5) \,a_0$.

The actual potential depth, $D_e$, relative to $^1S_0$-$^3P_1$ will be overestimated by the standard formula $\omega_e^2/4\omega_ex_e$ as the simple Morse model does not extrapolate well to the long-range. Instead, we determine $D_e$ by fitting the binding energies versus $v^\prime +\frac{1}{2}$ to Eq.~(\ref{eq:vibrotorsupp}), with $\alpha_e$, $\omega_ex_e$, and $\omega_e$ held fixed to the values above (see Fig.~\ref{fig:specparams}(c)). The vertical intercept of the quadratic fit will therefore be $2B_e-D_e$. Given knowledge of $B_e$, we find $D_e = 6387.89(11) \,\mathrm{cm}^{-1}$.

\section{Construction of Morse/Long-range Potentials\label{sec:mlr}}
The Morse/Long-range (MLR) potential as a function of internuclear separation $R$ has the form \begin{align}
  V_{\mathrm{MLR}} (R) = D_e \left[ 1 - \frac{u_{\mathrm{LR}}(R)}{u_{\mathrm{LR}}(R_e)} \mathrm{e}^{-\phi(R) \gamma_p(R)} \right] ^2, 
 \end{align}
 where 
 \begin{align}
 \label{eq:MLRformula}
  u_{\mathrm{LR}}(R) &= -\frac{C_3}{R^3} - \frac{C_6}{R^6} - \frac{C_8}{R^8} - \frac{C_{10}}{R^{10}}, 
 \\ \nonumber
 \phi(R) &= [1-\gamma_p^{\rm ref}(R)]\sum_{i=0}^N (\gamma^{\rm ref}_{q}(R))^i\phi_i + \gamma_p^{\rm ref}(R)\phi_{\infty},
 \\ \nonumber 
 \phi_{\infty} &= \ln \left(\frac{2 D_e}{u_{\text{LR}}(R_e)}\right),
 \\ \nonumber
 \gamma_p(R) &= \frac{R^p - R_e^p}{R^p + R_e^p}, 
 \\ \nonumber 
 \gamma_p^{\rm ref}(R) &= \frac{R^p - R_{\rm ref}^p}{R^p + R_{\rm ref}^p}.
 \end{align} As a starting point of the fitting procedure, we used the \textit{ab initio} results from Ref.~\cite{Skomorowski2012jcp} with the long-range coefficients fixed to the best available values deduced from experiments and \textit{ab initio} calculations: $C_3$ from Ref.~\cite{Borkowski2014}, $C_6$ and $C_8$ from Ref.~\cite{Porsev2014}, and $C_{10}$ from Ref.~\cite{Mitroy2010}. The well depth $D_e$ and the equilibrium distance $R_e$ of the $1_u$ potential were fixed at their empirical values found in this work. The $p, q$ and $R_{\text{ref}}$ parameters were chosen according to suggestions from Ref. \cite{LeRoy2009}. In particular, $p$ is an integer greater than $n_{\text{max}} - n_{\text{min}}$, where $n_{\text{min}} = 3$ and $n_{\text{max}} = 10$ are the lowest and the highest powers in the long range expansion $u_{LR}(R)$. Parameter $q$ is a smaller integer, typically $q = 2$-$4$ while $R_{\text{ref}}$ is chosen such as $R_{\text{ref}}/R_e = 1.1 - 1.5$. Introduction of additional parameters $q$ and $R_{\text{ref}}$ instead of simply using $p$ and $R_e$ leads to a better potential fit achieved with smaller number of terms in the expansion of $\phi(R)$ in Eq.~(\ref{eq:MLRformula}). Then the parameters $\phi_i$ of the $1_u$ potential were refitted to match the shape of the Morse potential well near the minimum. Finally, the parameters $\phi_0$, $\phi_1$, $\phi_2$  of the $1_u$ potential as well as the $C_6$ and $C_8$ long-range coefficients of both $0_u^+$ and $1_u$ potentials were fitted to a range of experimental $J^\prime=1$ binding energies given in Tab.~\ref{tab:bindingenergies}. A simultaneous fit of the long-range coefficients for both $0_u^+$ and $1_u$ potential curves was necessary to correctly describe the heavily Coriolis-mixed states $0_u^+(v=-6, J=1)$ and $1_u(v=-3, J=1)$. The resulting parameters are given in the upper half of Tab.~\ref{tab:mlr}. For convenience, the spectroscopic constants found in the preceding section are also listed in the lower half of Tab.~\ref{tab:mlr}.

\begin{table*}
 \caption{Binding energies for the bound states of $0_u^+$ and $1_u$, in units of MHz. Negative vibrational quantum numbers count down from the dissociation limit.  The $v^\prime$ assignments for the deeply bound $1_u$ states are explicitly described as belonging to either the MLR or the \textit{ab initio} (AI) potential. For the values obtained in this work, those of $1_u(v^\prime =23, 24, 26, J^\prime = 1, 3$) were determined via an optical frequency comb referenced to a NIST-traceable Rb standard, while the others were deduced with a commercial wavemeter. The uncertainty in the absolute binding energy of $X(v=6,J=2)$ dominates the uncertainty for the former, while the resolution of the wavemeter dominates for the latter. We use the $^1S_0$-$^3P_1$ intercombination-transition frequency from Ref. \cite{Ferrari2003}.}
 \label{tab:bindingenergies}
\begin{ruledtabular}
    \centering
    \begin{tabular}{lllllll}
State&$v^\prime$&$J^\prime$&\textit{Ab initio} (AI)&MLR&Experiment&Ref.\\
\colrule
$0_u^+$&-1&1&0.4438&0.4151&0.4653(45)&\cite{McDonald2017}\\
$0_u^+$&-2&1&23.8652&23.7511&23.9684(59)&\cite{McGuyer2015}\\
$0_u^+$&-3&1&223.967&221.643&222.161(35)&\cite{Zelevinsky2006}\\
$0_u^+$&-4&1&1,087.732&1,083.751&1,084.093(33)&\cite{Zelevinsky2006}\\
$0_u^+$&-5&1&3,421.12&3,463.63&3,463.28(33)&\cite{Zelevinsky2006}\\
$0_u^+$&-6&1&8,038.76&8,429.56&8,429.65(42)&\cite{Zelevinsky2006}\\
\colrule
$1_u$&-1&1&356.429&352.657&353.236(35)&\cite{Zelevinsky2006}\\
$1_u$&-2&1&2,686.905&2,684.103&2,683.722(32)&\cite{Zelevinsky2006}\\
$1_u$&-3&1&8,209.159&8,200.040&8,200.163(39)&\cite{Zelevinsky2006}\\
$1_u$&26 (MLR), 22 (AI)&1&129,004,636&129,586,955&129,584,370(2)&This work\\
$1_u$&25 (MLR), 21 (AI)&1&131,207,700&131,758,225&131,757,586(300)&This work\\
$1_u$&24 (MLR), 20 (AI)&1&133,424,436&133,942,696&133,943,644(2)&This work\\
$1_u$&23 (MLR), 19 (AI)&1&135,654,750&136,140,364&136,143,406(2)&This work\\
$1_u$&7 (MLR), 3 (AI)&1&173,112,867&173,081,203&173,083,976(300)&This work\\
$1_u$&6 (MLR), 2 (AI)&1&175,560,325&175,498,791&175,496,706(300)&This work\\
$1_u$&5 (MLR), 1 (AI)&1&178,019,781&177,928,703&177,926,824(300)&This work\\
\colrule
$1_u$&26 (MLR), 22 (AI)&3&128,998,655&129,580,953&129,578,356(2)&This work\\
$1_u$&24 (MLR), 20 (AI)&3&133,418,411&133,936,648&133,937,588(2)&This work\\
$1_u$&23 (MLR), 19 (AI)&3&135,648,703&136,134,293&136,137,328(2)&This work\\
    \end{tabular}
    \end{ruledtabular}
\end{table*}

\begin{table*}
 \caption{Morse/Long-range parameters of the $(1)0_u^+$ and $(1)1_u$ potentials in specified units. Also listed are the spectroscopic constants determined from the energies of deeply bound $1_u$ states.}
 \label{tab:mlr}
\begin{ruledtabular}
    \centering
    \begin{tabular}{lll}
&$0_u^+$&$1_u$\\
\colrule
$R_e$ ($a_0$) & 7.5443 & 7.9027 \\
$D_e$ (cm$^{-1}$) & 2784 & 6388 \\
$C_3$ ($E_h a_0^3$) & $1.5235661 \times 10^{-2}$ & $7.6178307\times 10^{-3}$\\
$C_6$ ($E_h a_0^6$) & $3.8947894\times 10^3$ & $4.0390241 \times 10^3$\\
$C_8$ ($E_h a_0^8$) & $4.5157846\times 10^5$ & $7.7660490 \times 10^5$\\
$C_{10}$ ($E_h a_0^{10}$) & $3.296\times 10^7$& $1.3253\times 10^8$ \\
$p$ & 9 & 9 \\
$q$ & 4 & 4 \\
$R_{\rm ref}$ ($a_0$) & 8.2987 & 8.6930 \\
$\phi_0$ & -0.63810976 & -1.2454828 \\
$\phi_1$ & 3.5917033 & -0.19418436 \\
$\phi_2$ & 7.7175691 & -1.8890781 \\
$\phi_3$ & 0.57800325 & -3.1121912 \\
$\phi_4$ & -29.3700406 & -6.0245946 \\
$\phi_5$ &  -23.080778 & -5.6268047 \\
$\phi_6$ & 54.044018 & -3.3425721 \\
$\phi_7$ & 91.862114 & -0.0028626398\\
$\phi_8$ & 35.061649 & 0\\
$\phi_9$ & -2.9283029 & 0\\
\colrule
$\omega_ex_e$ (cm$^{-1}$)&-& 0.21150(28) \\
$\omega_e$ (cm$^{-1}$)&-& 83.528(13)\\
$B_e$ (cm$^{-1}$)&-& 0.021933(3)\\
$\alpha_e$ (cm$^{-1}$)&-& 7.068(11)$\times 10^{-5}$
    \end{tabular}
    \end{ruledtabular}
\end{table*}

\end{document}